\documentclass[conference]{IEEEtran}
\IEEEoverridecommandlockouts
\AtBeginDocument{%
  \providecommand\BibTeX{{%
    Bib\TeX}}}

\usepackage{amsmath,amssymb,amsfonts}
\usepackage{graphicx}
\usepackage{textcomp}
\usepackage{booktabs} 
\usepackage{multirow}
\usepackage{hyperref}
\usepackage{url}
\usepackage{nicefrac}
\usepackage{microtype}
\usepackage{xcolor}
\usepackage{algorithm}
\usepackage{algpseudocode}
\usepackage{tikz}

\usetikzlibrary{positioning,arrows.meta,shapes.geometric,calc}

\usepackage{orcidlink}
\usepackage{comment}
\usepackage{braket}
\def\BibTeX{{\rm B\kern-.05em{\sc i\kern-.025em b}\kern-.08em
    T\kern-.1667em\lower.7ex\hbox{E}\kern-.125emX}}
\usepackage{cite}
\usepackage{pifont}
\usepackage{amsmath,amssymb,amsfonts}
\usepackage{graphicx}
\usepackage{textcomp}
\usepackage{amsmath}
\usepackage{booktabs} 
\usepackage{color}
\usepackage{multirow}
\usepackage{colortbl, xcolor}

\usepackage{hyperref}       
\usepackage{url}            
\usepackage{booktabs}       
\usepackage{amsfonts}       
\usepackage{nicefrac}       
\usepackage{microtype}      
\usepackage{xcolor}         
\usepackage{algorithm}
\usepackage{algpseudocode}

\usepackage[skip=3pt]{caption}
\usepackage{amsmath}
\usepackage{tikz}

\usepackage{orcidlink}
\usepackage{comment}
\usepackage{amssymb}
\usepackage{braket}
\usepackage{xcolor}
\usepackage[table]{xcolor}

\definecolor{classicalgray}{RGB}{235,235,235}
\definecolor{quantumblue}{RGB}{226,240,255}
\definecolor{metagreen}{RGB}{226,245,230}
\usepackage{enumitem}
\setlist[itemize]{leftmargin=*}%
\setlist[enumerate]{leftmargin=*}%

\begin{document}
\title{Meta-Quantum Ensemble Framework for Robust Network Intrusion Detection}

\author{\IEEEauthorblockN{Ritvik Bhatnagar\orcidlink{0009-0008-5818-2309}\textsuperscript{1}, Nouhaila Innan\orcidlink{0000-0002-1014-3457}\textsuperscript{2,3}, Angel Arul Jothi J.\orcidlink{0000-0002-1773-8779}\textsuperscript{1}, Muhammad Shafique\orcidlink{0000-0002-2607-8135}\textsuperscript{2,3}\\
\IEEEauthorblockA{
\textsuperscript{1}Department of Computer Science, Birla Institute of Technology and Science, Pilani,\\\ Dubai Campus, Dubai International Academic City, Dubai, UAE\\
\textsuperscript{2}eBRAIN Lab, Division of Engineering, New York University Abu Dhabi (NYUAD), Abu Dhabi, UAE\\
\textsuperscript{3}Center for Quantum and Topological Systems (CQTS), NYUAD Research Institute, NYUAD, Abu Dhabi, UAE\\
Emails: \{f20220031,angeljothi\}@dubai.bits-pilani.ac.in, \{nouhaila.innan, muhammad.shafique\}@nyu.edu\\
}}}
\maketitle
\begin{abstract}
Intrusion Detection Systems (IDSs) must maintain high detection sensitivity while operating under strict false-positive constraints, a challenge intensified by class imbalance and heterogeneous IoT traffic. This work investigates whether heterogeneous quantum learners can provide useful and non-redundant decision information for IDS tasks. We study Quantum Support Vector Machines (QSVMs) and Quantum Neural Networks (QNNs), which rely on different learning mechanisms and exhibit distinct prediction behaviors. To combine these models, we propose the System-Level Meta-Quantum Ensemble (MQE), a hybrid quantum--classical framework that fuses QSVM and QNN outputs using a Random Forest meta-learner. The meta-learner captures agreement and disagreement patterns between the quantum branches to improve prediction stability and detection performance. Experiments on TON IoT and CICIDS2017 show that MQE improves selected performance, low-FPR, and reliability metrics over several standalone quantum learners, with gains depending on the dataset, metric, and fusion representation. The results highlight meta-level fusion as a practical strategy for building more reliable QML-based IDS pipelines.
\end{abstract}

\section{Introduction}

\begin{figure}[htbp]
    \centering
    \includegraphics[width=1\linewidth]{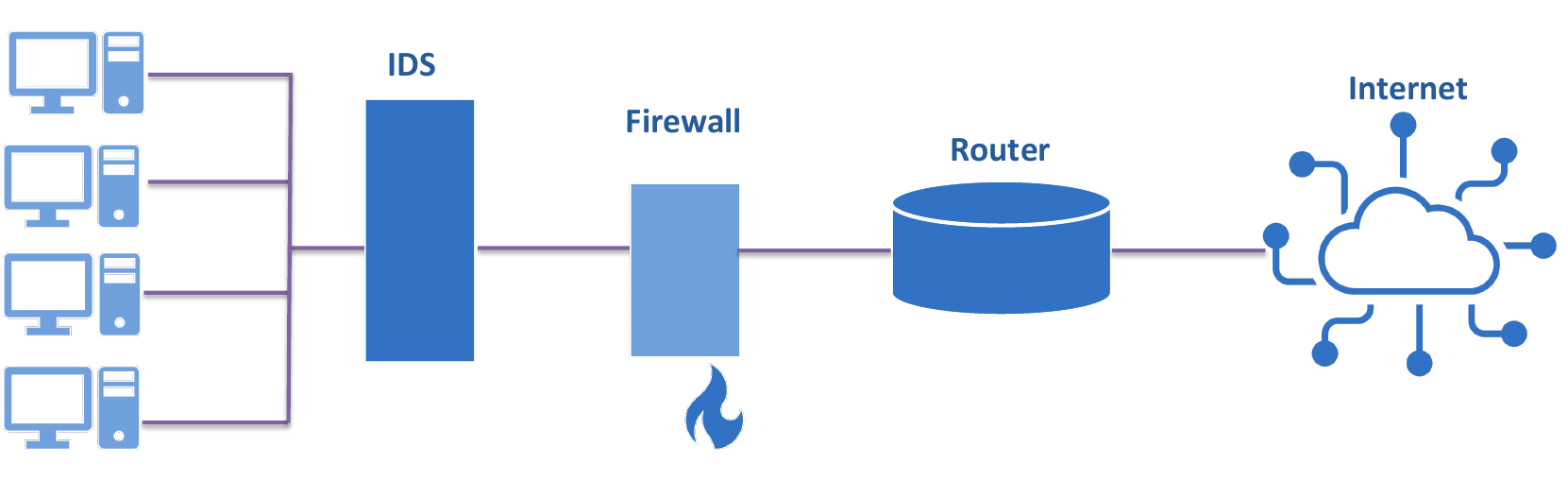}
    \caption{Typical network architecture with IDS deployment.}
    \label{fig:ids}
\end{figure}

Modern network infrastructures, particularly those deployed in Internet of Things (IoT) environments, are increasingly vulnerable to sophisticated and rapidly evolving cyber threats. As these threats become more diverse and dynamic, Intrusion Detection Systems (IDS) (see Fig.~\ref{fig:ids}) play an essential defensive role by continuously monitoring network traffic and system activity for signs of malicious behavior \cite{khraisat2019survey,bilot2025sometimes}. Despite their importance, achieving high detection rates while maintaining extremely low false-positive rates remains an open challenge. This difficulty is further compounded by three factors common in real-world IoT settings: (i) severe class imbalance between benign and malicious traffic, (ii) non-stationary data distributions caused by concept drift, and (iii) the need for scalable and adaptive detection under resource constraints.

Classical machine learning models, including advanced ensembles such as XGBoost and deep neural architectures~\cite{sensors2023, sohail2023deep}, already achieve strong performance in intrusion detection and remain the state of the art for many operational systems. The goal of this work is not to replace these mature classical approaches, but to explore whether quantum machine learning (QML) can provide promising modeling benefits. Quantum models use quantum kernels and variational circuits to embed data into high-dimensional Hilbert spaces \cite{schuld2021supervised,biamonte2017quantum}, potentially capturing structure that classical methods may overlook under complex or evolving conditions. 

Among QML approaches, two models are particularly relevant for IDS applications: Quantum Support Vector Machines (QSVMs) \cite{rebentrost2014quantum,innan2023enhancing}, which exploit quantum kernels to enhance feature separability, and Quantum Neural Networks (QNNs) \cite{schuld2014quest,innan2025next}, which employ variational circuits to learn expressive nonlinear boundaries. Our preliminary findings, along with those from prior studies \cite{larocca2023theory,garcia2024effects,njiki2026robustness,vyskubov2026scaling}, show that these models exhibit interesting behaviors: QSVMs offer stable and well-defined decision margins but limited adaptivity, whereas QNNs are more flexible but prone to noise sensitivity and overfitting. These contrasting properties highlight the value of combining them.

In parallel, existing research in classical IDS has demonstrated that meta-learning and ensemble fusion techniques significantly improve robustness and detection performance \cite{sohail2023deep,mamatha2025development,mahmoud2025dsem,pramanick2025leveraging}, especially under imbalance and distributional variability. This raises a natural question: \emph{Can similar meta-ensemble strategies also enhance quantum machine learning models, whose strengths and weaknesses differ in important ways from classical learners?} Motivated by this, we explore whether ensemble integration can amplify the complementary capabilities of QSVMs and QNNs.

To this end, we propose a \textbf{System-Level Meta-Quantum Ensemble (MQE)} framework that fuses the outputs of QSVM and QNN models using a classical Random Forest meta-learner. The framework is designed to exploit the distinct inductive biases of the two quantum learners: the kernel-based geometric decision structure of QSVM and the trainable nonlinear representation learned by QNN. By operating on their output scores, the meta-learner captures agreement and disagreement patterns between the two branches, leading to more reliable predictions under class imbalance and heterogeneous traffic conditions.

The key contributions of this work are summarized below:
\begin{itemize}
    \item We introduce a System-Level Meta-Quantum Ensemble framework that integrates QSVM and QNN within a hybrid quantum--classical IDS pipeline.
    \item We develop a Random Forest-based meta-fusion mechanism that combines the distinct prediction behaviors of the two quantum learners and improves performance over several standalone quantum branches.
    \item We provide a detailed evaluation on TON~IoT and CICIDS2017, including standard classification metrics, low-FPR detection behavior, meta-fusion analysis, noise robustness, and calibration reliability.
\end{itemize}
The rest of the paper is structured as follows: Section~\ref{sec:back} discusses related work. Section~\ref{sec:methodology} details the proposed methodology. Section~\ref{sec:results} presents experimental results and analysis, and Section~\ref{sec:conclusion} concludes the paper.
\section{Background and Related Work}
\label{sec:back}

Classical intrusion detection research has produced a wide range of solutions, including deep neural networks, boosted ensembles, and meta-learning architectures, all of which have demonstrated strong performance in complex network environments. However, since the focus of this work is on quantum approaches, we summarize here the growing body of research applying QML to IDS and related cybersecurity tasks. Recent studies have explored hybrid quantum–classical neural models for detecting network attack traffic, showing that variational circuits can improve feature discrimination in limited-data regimes~\cite{wang2025network}. Other works, such as QuantumNetSec and similar QSVM/VQC-based IDS frameworks, report that quantum kernels and parameterized circuits can achieve competitive or even superior detection performance on IoT-network datasets~\cite{abreu2025quantumnetsec, kumar2025quids}. Further developments include quantum autoencoder-based anomaly detectors~\cite{hdaib2024quantum}, and quantum-enhanced threat analysis frameworks~\cite{sahoo2025qcshield,chaudhary2025towards}. Collectively, these studies show that QML is an emerging and promising direction for IDS, capable of employing quantum feature spaces to model complex attack patterns.

Beyond IDS-specific research, the broader QML community has developed techniques directly relevant to meta-modeling and multi-model coordination. Quantum meta-learning frameworks, such as quantum model-agnostic meta-learning and quantum multi-agent adaptive learning~\cite{yun2023quantum,lee2025q}, demonstrate that meta-level optimization, rapid adaptation, and coordinated decision-making are feasible within quantum settings. However, these approaches have not yet been applied to intrusion detection, nor have they been investigated as mechanisms for combining heterogeneous quantum learners.

Despite these advances, current QML–IDS studies share a common limitation: they evaluate individual quantum models in isolation. QSVMs, QNNs, QCNNs, and VQCs are typically compared as standalone classifiers rather than integrated into a unified decision structure. Moreover, very few works address low–false-positive constraints, robustness under quantum noise, or the potential benefits of structured fusion mechanisms for quantum learners. To the best of our knowledge, no prior research has proposed a hybrid quantum–classical \emph{meta-ensemble} that fuses distinct quantum models within a single intrusion detection pipeline. This gap motivates our exploration of whether meta-modeling principles, successful in classical IDS, can similarly enhance the performance, stability, and reliability of quantum intrusion detection systems.

\section{Methodology}
\label{sec:methodology}

\begin{figure*}[htbp]
    \centering
    \includegraphics[width=1\linewidth]{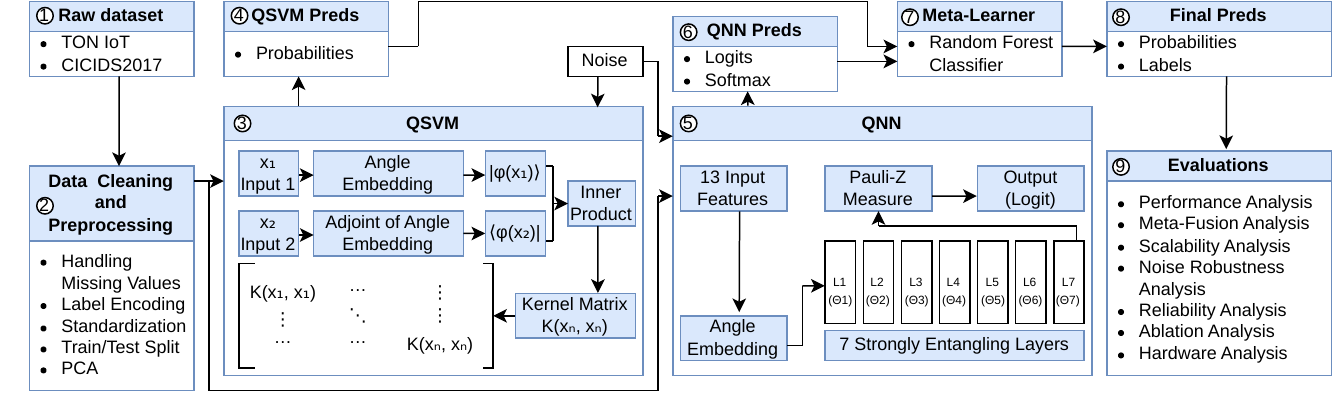}
    \caption{System architecture illustrating data preprocessing, quantum feature encoding, QNN and QSVM branches, and the final meta-learning fusion stage.}
    \label{fig:arch_diag}
\end{figure*}

The MQE framework integrates two quantum learners, QSVM and QNN, with a classical Random Forest meta-model (see Fig.~\ref{fig:arch_diag}). After preprocessing and angle embedding, the QSVM branch computes a quantum kernel matrix for classical SVM classification, while the QNN branch uses a variational circuit to learn nonlinear boundaries. The meta-model fuses the probabilistic outputs of both learners, resolving disagreements and improving robustness. In essence, QSVM offers stable margins, QNN captures nonlinear drift, and the meta-learner combines their strengths to maintain a high true-positive rate (TPR) while keeping the false-positive rate (FPR) low.

\subsection{Data Cleaning and Preprocessing}

The raw datasets contain heterogeneous numerical and categorical attributes across different scales. To ensure compatibility with both quantum and classical components, we apply a consistent preprocessing pipeline: categorical attributes are label-encoded, missing values are removed, and all features are standardized to zero mean and unit variance. 

To reduce feature dimensionality and control the number of qubits required, Principal Component Analysis (PCA) is applied, retaining the most informative components while reducing redundancy and improving circuit efficiency. The resulting feature vectors serve as inputs to both QSVM and QNN learners.

\subsection{Quantum Feature Embedding}

Classical feature vectors are transformed into quantum states via angle embedding, chosen for its compatibility with variational circuits and its adaptability to the selected datasets.
Each feature $x_i$ is encoded as a rotation around the $Y$-axis applied to a single qubit:
$|\psi(\mathbf{x})\rangle = \bigotimes_{i=1}^{n} R_Y(x_i)\,|0\rangle,$
where $n$ is the number of qubits, this embedding preserves relative feature magnitudes and maps the data into a high-dimensional Hilbert space, enabling downstream models to leverage entanglement and superposition for richer representations.

\subsection{QNN Model}
\begin{figure}[htbp]
    \centering
    \includegraphics[width=0.75\linewidth]{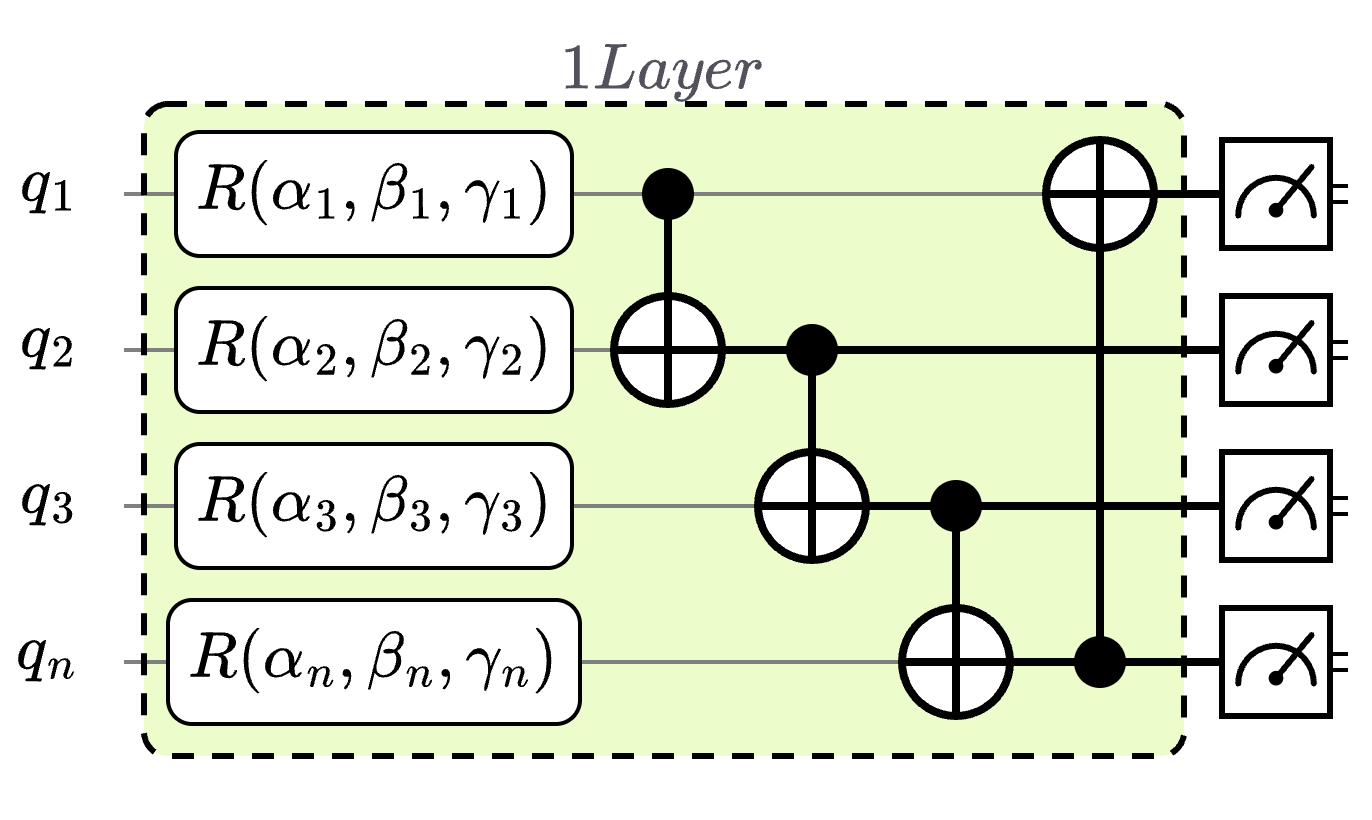}
\caption{Single-layer QNN circuit based on a strongly entangling structure. Each qubit is processed by a parameterized rotation gate \(R(\alpha_i,\beta_i,\gamma_i)\), followed by CNOT-based entangling operations and measurement.}
    \label{qnn}
\end{figure}
The QNN branch employs a Variational Quantum Circuit (VQC) constructed using Strongly Entangling Layers (SELs), which consist of parameterized single-qubit rotations followed by CNOT-based entanglement arranged in a ring topology (see Fig. \ref{qnn}). The full ansatz is
\begin{equation}
U(\boldsymbol{\theta}) 
= \prod_{l=1}^{L} 
\left(
\prod_{i=1}^{n} 
R_Z(\theta_{l,i}^{(3)})
R_Y(\theta_{l,i}^{(2)})
R_X(\theta_{l,i}^{(1)})
\right)
U_{\text{ent}}^{(l)},
\end{equation}
where $L$ is the circuit depth and $\boldsymbol{\theta}$ denotes trainable parameters. Expressivity increases with depth, analogous to deep classical neural networks.

After encoding and applying $U(\boldsymbol{\theta})$, the circuit is measured in the Pauli-$Z$ basis:
\begin{equation}
z_i = \langle \psi(\mathbf{x},\boldsymbol{\theta}) | Z_i | \psi(\mathbf{x},\boldsymbol{\theta})\rangle,
\quad i = 1,\dots,n.
\end{equation}
The resulting expectation values form the input to a classical dense layer for binary classification. Parameters are optimized using a classical gradient-based optimizer until convergence.
\subsection{QSVM Model}

\begin{figure}[htbp]
    \centering
    \includegraphics[width=0.75\columnwidth]{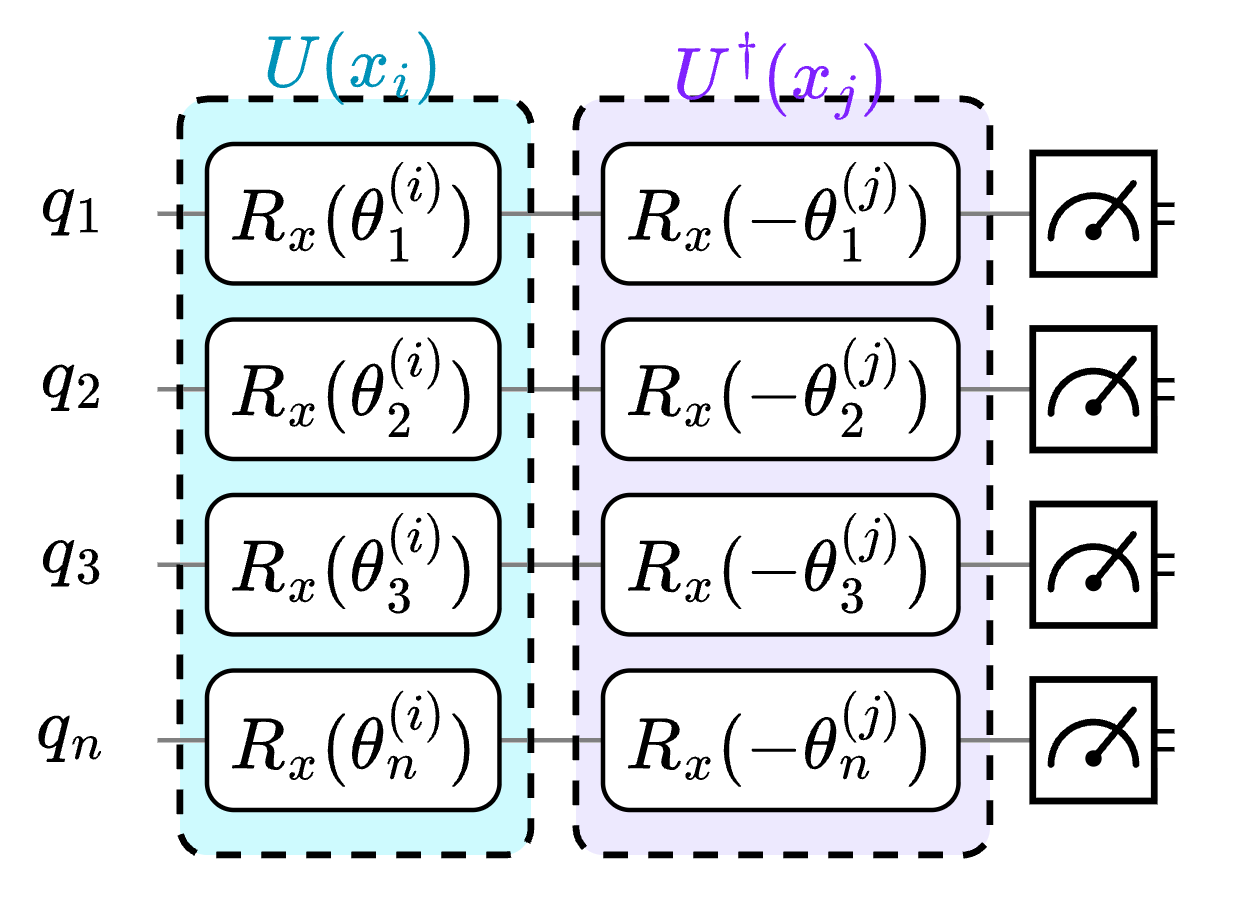}
   \caption{QSVM kernel circuit representation, where the input sample \(x_i\) is encoded through the feature map \(U(x_i)\), followed by the adjoint feature map \(U^\dagger(x_j)\). The kernel value \(K(x_i,x_j)\) is estimated from the probability of measuring the all-zero state after applying \(U^\dagger(x_j)U(x_i)\).}
    \label{fig:qsvm_circuit}
\end{figure}
The QSVM branch uses a quantum kernel method to compute sample similarities in a quantum-induced feature space (see Fig. \ref{fig:qsvm_circuit}). For two inputs $x_i$ and $x_j$, the kernel value is defined as the fidelity:
\begin{equation}
K(x_i, x_j) = |\langle \psi(x_i) | \psi(x_j) \rangle|^2.
\end{equation}
The kernel circuit embeds one sample, applies the adjoint embedding of the other, and measures the overlap. This produces a normalized and symmetric similarity measure that populates the Quantum Kernel Matrix:
\begin{equation}
K_{ij} = K(x_i, x_j).
\end{equation}
The resulting Hermitian, positive semi-definite matrix is used by a classical SVM to determine the optimal separating hyperplane:
\begin{equation}
f(x) = \text{sign}\!\left(\sum_{i=1}^{m} \alpha_i y_i K(x_i, x) + b \right).
\end{equation}
This hybrid design captures complex geometric relationships while leveraging classical SVM optimization.

\subsection{Random Forest Meta-Model}

The Random Forest meta-model serves as the fusion layer combining outputs from the QNN and QSVM branches. For each input $i$, the concatenated meta-feature is
\begin{equation}
\mathbf{p}_i = \left[p^{(\text{QNN})}_i,\; p^{(\text{QSVM})}_i\right],
\end{equation}
where $p$ denotes predicted attack probability. The RF consists of an ensemble of decision trees, each trained on random subsets of data and features. Final predictions are obtained via majority voting:
\begin{equation}
\hat{y}_i = \operatorname{mode}\{t_j(\mathbf{p}_i)\}_{j=1}^{T}.
\end{equation}
The meta-model exploits complementary strengths, QSVM’s stable margins and QNN’s flexible nonlinear representations, yielding improved robustness and reduced variance.

\subsection{Evaluation Pipeline}

The evaluation pipeline assesses the individual components (QSVM and QNN) as well as the full ensemble across several dimensions: accuracy, precision, recall, F1-score, area under the precision–recall curve (AUPRC), and TPR at low FPR thresholds. 

Beyond standard metrics, the evaluation pipeline includes additional analyses to assess robustness and reliability. We test noise resilience by simulating common quantum noise channels, such as depolarizing and amplitude-damping noise, and evaluate performance under each noise condition. We also evaluate calibration using the Brier Score,
\begin{equation}
\text{Brier Score} = \frac{1}{N} \sum_{i=1}^{N} (p_i - y_i)^2,
\end{equation}
where \(p_i\) and \(y_i\) denote the predicted probability and true label, and the Expected Calibration Error (ECE),
\begin{equation}
\text{ECE} = \sum_{m=1}^{M} \frac{n_m}{N} \left| \text{acc}(m) - \text{conf}(m) \right|,
\end{equation}
where \(n_m\), \(\text{acc}(m)\), and \(\text{conf}(m)\) represent the bin count, accuracy, and mean confidence, respectively. These analyses provide a compact yet comprehensive view of robustness, calibration, and NISQ feasibility.

\section{Results and Discussion \label{sec:results}}
\subsection{Experimental Setup}

We evaluate the proposed MQE framework on two benchmark intrusion detection datasets: \textbf{CICIDS2017 (D1)}~\cite{dataset_d1} and \textbf{TON IoT (D2)}~\cite{dataset_d2}. CICIDS2017 contains approximately 2.8 million labeled network-flow samples with 79 features and covers multiple attack categories under realistic traffic conditions. TON~IoT contains 211{,}042 samples with 44 features collected from network traffic, telemetry, and system logs, providing a heterogeneous IoT-oriented evaluation setting.

For both datasets, categorical attributes are label-encoded, missing values are removed, and numerical features are standardized using statistics computed only from the training split. PCA is then applied to obtain a 13-dimensional representation, matching the 13-qubit quantum input size used by both the QSVM and QNN branches. The same preprocessed feature representation is used across all quantum models and meta-fusion variants to ensure a controlled comparison.

All quantum experiments are executed using the \texttt{PennyLane} \texttt{lightning.qubit} simulator backend. The full pipeline is run on a machine with an Intel Core i7-13620 CPU, 16\,GB RAM, and an NVIDIA RTX~4060 GPU with 8\,GB VRAM under Ubuntu~WSL. Table~\ref{tab:experimental_settings} summarizes the main hyperparameters used for the quantum models and the Random Forest meta-learner.

The evaluation includes accuracy, F1-score, AUPRC, ROC-AUC, and true-positive rate (TPR) under strict false-positive rate thresholds of 0.1\% and 1\%. These low-FPR metrics are included because intrusion detection systems often require high detection sensitivity while limiting false alarms in operational settings.

\begin{table}[htbp]
\centering
\caption{Hyperparameter settings for the quantum learners and Random Forest meta-learner.}
\label{tab:experimental_settings}
\begin{tabular}{lcc}
\toprule
\textbf{Category} & \textbf{Parameter} & \textbf{Description} \\
\midrule
\multirow{8}{*}{Quantum Models}
 & Number of Qubits & 13 \\
 & QNN Layers & 7 \\
 & QNN Optimizer & Adam \\
 & QNN Learning Rate & 0.01 \\
 & QNN Batch Size & 32 \\
 & QNN Epochs & 100 \\
 & QNN Loss Function & Cross-Entropy \\
 & QSVM Regularization & 10 \\
 & Train/Test Split & 80\% / 20\% \\
\midrule
\multirow{3}{*}{Meta-Model} 
 & Number of Trees & 100 \\
 & Max Depth & Unrestricted \\
 & Criterion & Gini Impurity \\
\bottomrule
\end{tabular}%
\end{table}

\subsection{Performance Analysis}

We first evaluate the overall detection performance of the proposed MQE framework before analyzing the behavior of its individual components. This analysis assesses whether the meta-ensemble can maintain high detection quality across two intrusion detection settings with different feature distributions, class imbalance levels, and traffic characteristics.

As shown in Table~\ref{tab:performance}, MQE achieves strong performance on both datasets, although the difficulty level differs between D1 and D2. On CICIDS2017 (D1), the framework reaches 97.14\% accuracy and an F1-score of 81.48\%. This result indicates reliable detection under a more challenging setting, where class imbalance and heterogeneous attack patterns make positive-class detection harder. On TON~IoT (D2), MQE achieves 98.57\% accuracy and a 99.05\% F1-score, showing stronger class separability and more stable classification behavior. The AUPRC values further support this observation, with 86.22\% on D1 and 99.55\% on D2. At strict false-positive constraints, the model obtains TPR values of 40.74\% and 70.37\% on D1 at 0.1\% and 1\% FPR, respectively, while D2 reaches 7.89\% and 78.92\%. These results suggest that MQE provides robust detection performance, but also that low-FPR sensitivity remains more dataset-dependent than standard accuracy or F1-score.

\begin{table}[htpb]
\centering
\caption{Performance metrics of the proposed MQE framework on D1 and D2.}
\label{tab:performance}
\resizebox{\columnwidth}{!}{%
\begin{tabular}{lccccc}
\toprule
 & F1 (\%)& Acc (\%)& AUPRC (\%)& TPR@0.1\%FPR (\%)& TPR@1\%FPR (\%)\\
\midrule
D1 & 81.48 & 97.14 & 86.22 & 40.74 & 70.37  \\
D2 & 99.05 & 98.57 & 99.55 & 7.89 & 78.92 \\
\bottomrule
\end{tabular}
}
\end{table}
\begin{figure}[htbp]
    \centering
    \includegraphics[width=1\linewidth]{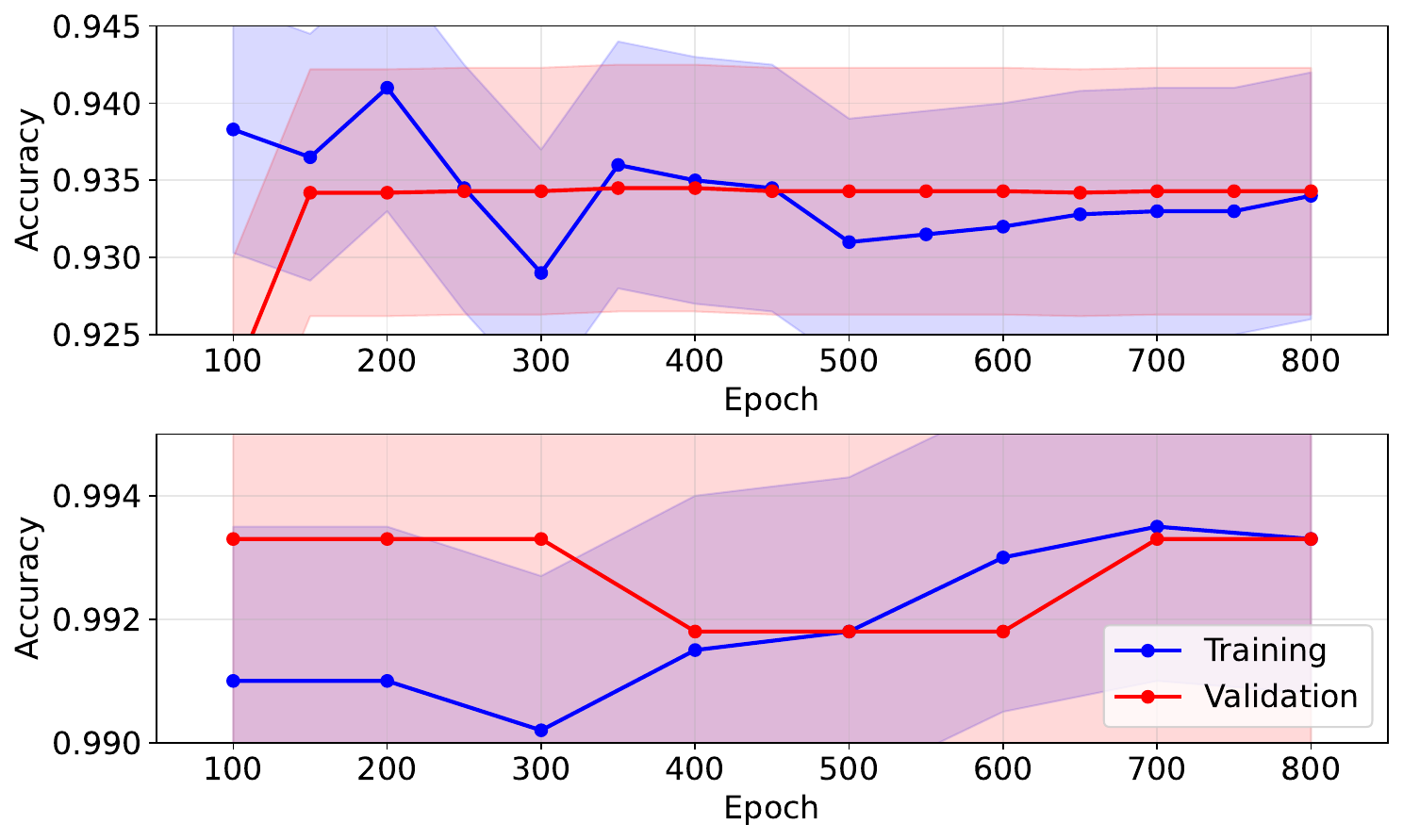}
    \caption{Learning curves for D1 (top) and D2 (bottom), showing smooth and stable convergence.}
    \label{fig:learning_cruve}
\end{figure}

To further examine learning behavior, Fig.~\ref{fig:learning_cruve} shows the training and validation accuracy curves for both datasets. The curves indicate stable convergence without a clear widening gap between training and validation accuracy, suggesting limited overfitting. For D1, the accuracy stabilizes near 97\%, with visible fluctuations that likely reflect the stronger class imbalance and more heterogeneous traffic patterns in CICIDS2017. In contrast, D2 exhibits smoother convergence and reaches approximately 98.6\% accuracy, indicating a more stable optimization process and clearer feature separation. In some epochs, validation accuracy slightly exceeds training accuracy; this behavior can occur when the validation split contains less difficult samples or a different class composition, and does not by itself indicate overfitting.

\begin{figure}
    \centering
    \includegraphics[width=1\linewidth]{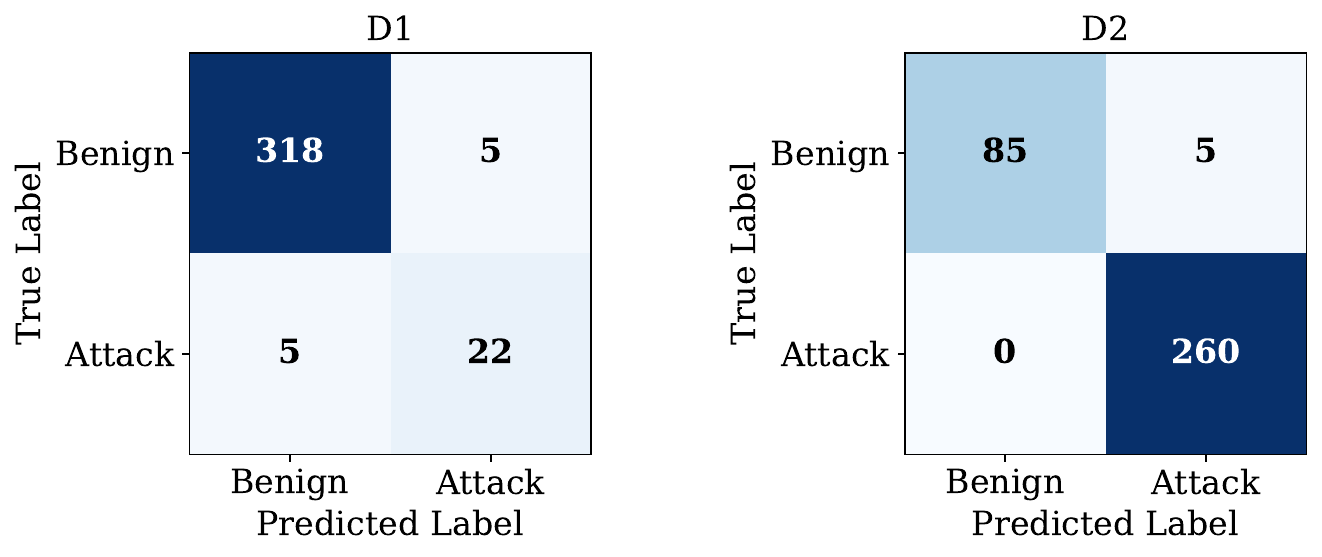}
    \caption{Confusion matrices for D1 and D2 demonstrating strong classification performance with minimal errors.}
    \label{fig:cm}
\end{figure}
The confusion matrices in Fig.~\ref{fig:cm} further clarify the classification behavior of MQE. For D1, the model correctly classifies 318 benign samples and 22 attack samples, with 5 false positives and 5 false negatives. This balanced error pattern indicates that the model maintains reliable benign-traffic recognition while still detecting a meaningful portion of the minority attack class. For D2, the model correctly identifies 85 benign samples and 260 attack samples, with only 5 false positives and no false negatives. This confirms the stronger detection behavior observed in Table~\ref{tab:performance}, especially for attack identification on TON~IoT.

These results provide the basis for the following meta-fusion analysis, where we examine whether the improved behavior comes from the interaction between the QSVM and QNN branches rather than from a single quantum learner alone.

\subsection{Meta-Fusion Analysis}

To understand why the meta-model improves upon the individual quantum learners, we first analyze the degree to which QSVM and QNN make different prediction errors. Their complementarity is quantified through the \textbf{disagreement rate}, defined as the fraction of samples for which the QSVM and QNN predicted labels differ under a decision threshold of 0.5. We also report the sizes of the individual error sets, $E_{\text{QSVM}}$ and $E_{\text{QNN}}$, as well as their intersection, which corresponds to samples misclassified by both models.

\begin{table}[htpb]
\centering
\caption{Error Sets and Disagreement Rate}
\label{tab:error_rate}
\resizebox{\columnwidth}{!}{%
\begin{tabular}{lcccc}
\toprule
 & QSVM & QNN & Both & Disagreement Rate (\%)\\
\midrule
Errors in D1 & 97 & 62 & 46 & 6.7\\
Errors in D2 & 28 & 7 & 4 & 2.7\\
\bottomrule
\end{tabular}
}
\end{table}

\begin{figure}[htbp]
    \centering
    \includegraphics[width=1\linewidth]{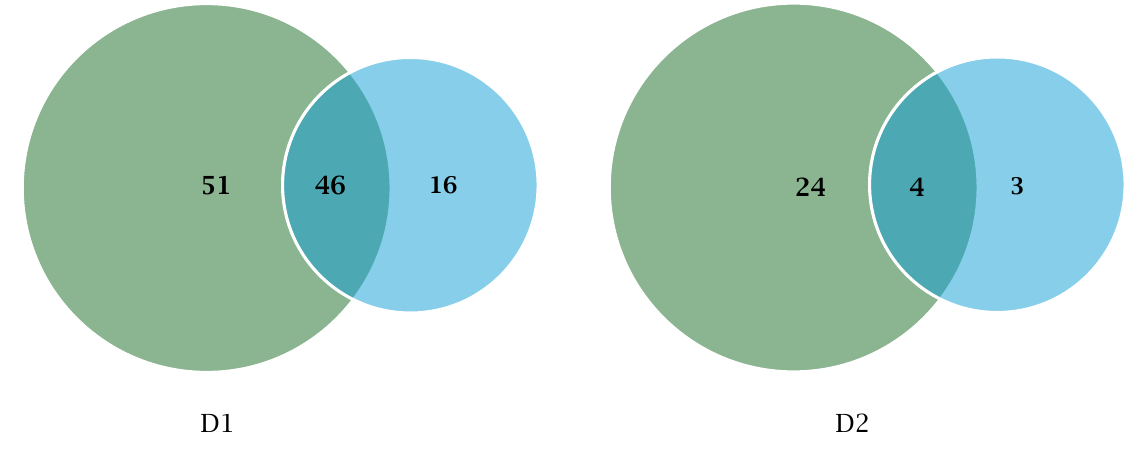}
    \caption{Venn diagrams showing error overlap between QSVM and QNN for D1 and D2.}
    \label{fig:error_venns}
\end{figure}

As shown in Table~\ref{tab:error_rate} and Fig.~\ref{fig:error_venns}, the relationship between the two learners differs across datasets. On D1, both models make more errors, and the shared error set is relatively large, indicating that CICIDS2017 contains a harder subset of samples that remains challenging for both quantum learners. On D2, the number of errors is much smaller and the overlap is more limited, suggesting that QSVM and QNN fail on more distinct instances. This lower overlap indicates stronger complementarity, which is beneficial for the meta-learner because it can exploit disagreements between the base models rather than inheriting the same errors from both.
To further examine how the fusion stage uses this information, we compare three meta-feature extraction schemes: \emph{hard}, \emph{margin}, and \emph{probabilistic} fusion. These schemes provide the Random Forest meta-model with different forms of information, ranging from discrete decisions to richer confidence-based signals. The results are reported in Table~\ref{tab:meta_feature}.

\begin{table}[h]
\centering
\caption{Meta-Feature Fusion Results}
\label{tab:meta_feature}
\resizebox{\columnwidth}{!}{%
\begin{tabular}{lccll}
\toprule
 & F1 (D2) (\%)& TPR@1\%FPR (D2) (\%)& F1 (D1) (\%)& TPR@1\%FPR (D1) (\%)\\
\midrule
Hard  & 99.43& 89.31& 69.77& 62.99\\
Margin & 99.04& 77.54& 78.43& 74.07\\
Prob  & 99.05& 78.92& 81.48& 70.37\\
\bottomrule
\end{tabular}
}
\end{table}

The three fusion schemes exhibit distinct behaviors across the two datasets. For D2, hard fusion achieves the strongest results, with an F1-score of 99.43\% and a TPR@1\%FPR of 89.31\%. This suggests that, on TON~IoT, the base learners already produce sufficiently reliable class decisions, allowing the meta-model to benefit most from their discrete outputs. In contrast, margin and probabilistic fusion produce nearly identical results on D2, with F1-scores close to 99.05\% and TPR@1\%FPR values around 78\%, indicating that additional confidence information brings limited extra benefit in this setting.

For D1, the pattern differs. Probabilistic fusion achieves the best F1-score at 81.48\%, while margin fusion provides the highest TPR@1\%FPR at 74.07\%. Hard fusion performs noticeably worse, with an F1-score of 69.77\%. This indicates that, for CICIDS2017, the discrete outputs alone do not carry enough information for effective fusion. Instead, the continuous scores provided by margin and probability-based representations offer the meta-model a more informative view of uncertainty, which is especially useful under stronger class imbalance and greater overlap between benign and attack samples.

These results show that the value of fusion depends not only on the complementarity between QSVM and QNN, but also on how their outputs are represented. On D2, where the base learners are more separable and their shared error set is small, hard fusion is sufficient and even preferable. On D1, where the task is more difficult and the shared error region is larger, richer confidence-based meta-features allow the Random Forest to better distinguish borderline cases. This confirms that the proposed MQE framework does not depend on a single fixed fusion rule; instead, it can adapt to dataset-specific characteristics and extract useful information from different output formats. These findings support the role of meta-fusion as a key component in improving robustness and predictive quality over standalone quantum learners.
\subsection{Noise Robustness Analysis}

\begin{figure}[htbp]
    \centering
    \includegraphics[width=1\linewidth]{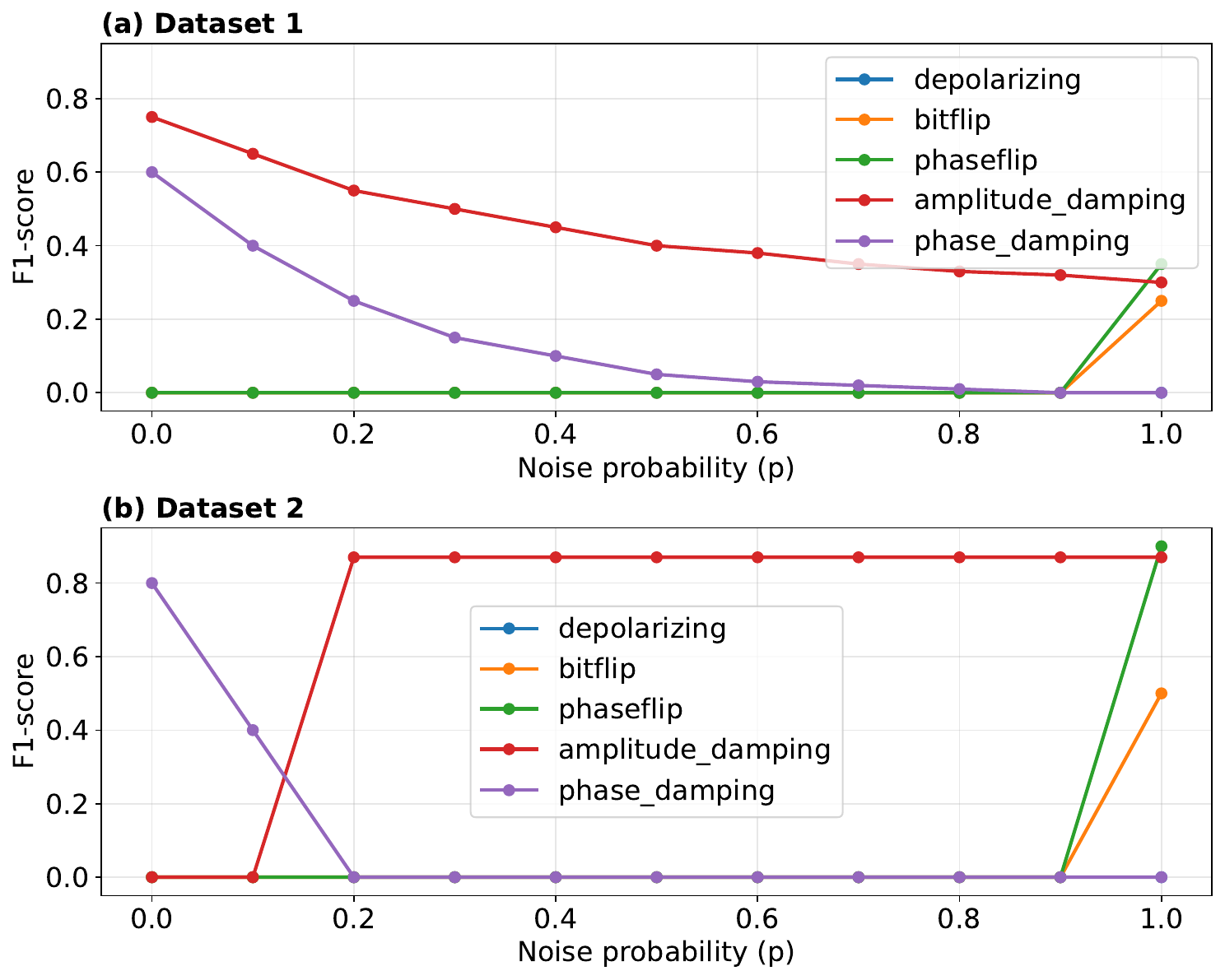}
    \caption{Noise robustness of the proposed MQE framework under different quantum noise types for datasets D1 and D2.}
    \label{fig:noise}
\end{figure}

To assess the resilience of the proposed MQE framework under imperfect quantum execution, we evaluate its performance under five common noise models: amplitude damping, bit flip, phase flip, phase damping, and depolarizing noise. Fig.~\ref{fig:noise} shows how the F1-score changes as the noise probability increases for both datasets.

The results indicate that MQE preserves reasonable robustness under low-to-moderate noise levels, although the sensitivity depends on both the dataset and the noise type. For D2, the framework remains comparatively stable under amplitude damping, with only limited degradation across much of the noise range. For D1, the same noise type leads to a more visible but still gradual decrease in performance, indicating that the CICIDS2017 setting is more sensitive to perturbations in the quantum computations.

For the other noise models, namely bit flip, phase flip, phase damping, and depolarizing noise, performance generally decreases as the noise probability increases. This trend is consistent with the expected impact of decoherence and stochastic gate disturbances on the encoded quantum states and the resulting model outputs. The degradation is more pronounced at larger noise levels, where the information carried by the quantum representations becomes increasingly corrupted.

At very high noise probabilities, some curves exhibit small upward fluctuations. These should not be interpreted as true performance recovery. Rather, they reflect unstable prediction behavior when the circuit outputs become highly randomized under extreme noise. In such regimes, the measured performance may vary irregularly because the learned quantum representation is no longer reliably preserved.

These results show that the proposed MQE framework remains reasonably robust in the presence of moderate noise, but its performance degrades as noise becomes more severe. This behavior is expected for NISQ-oriented quantum models and suggests that future implementations may benefit from error-mitigation techniques, noise-aware training, or more noise-tolerant circuit designs.
\subsection{Calibration and Reliability Analysis}

\begin{figure}[htbp]
    \centering
    \includegraphics[width=1\linewidth]{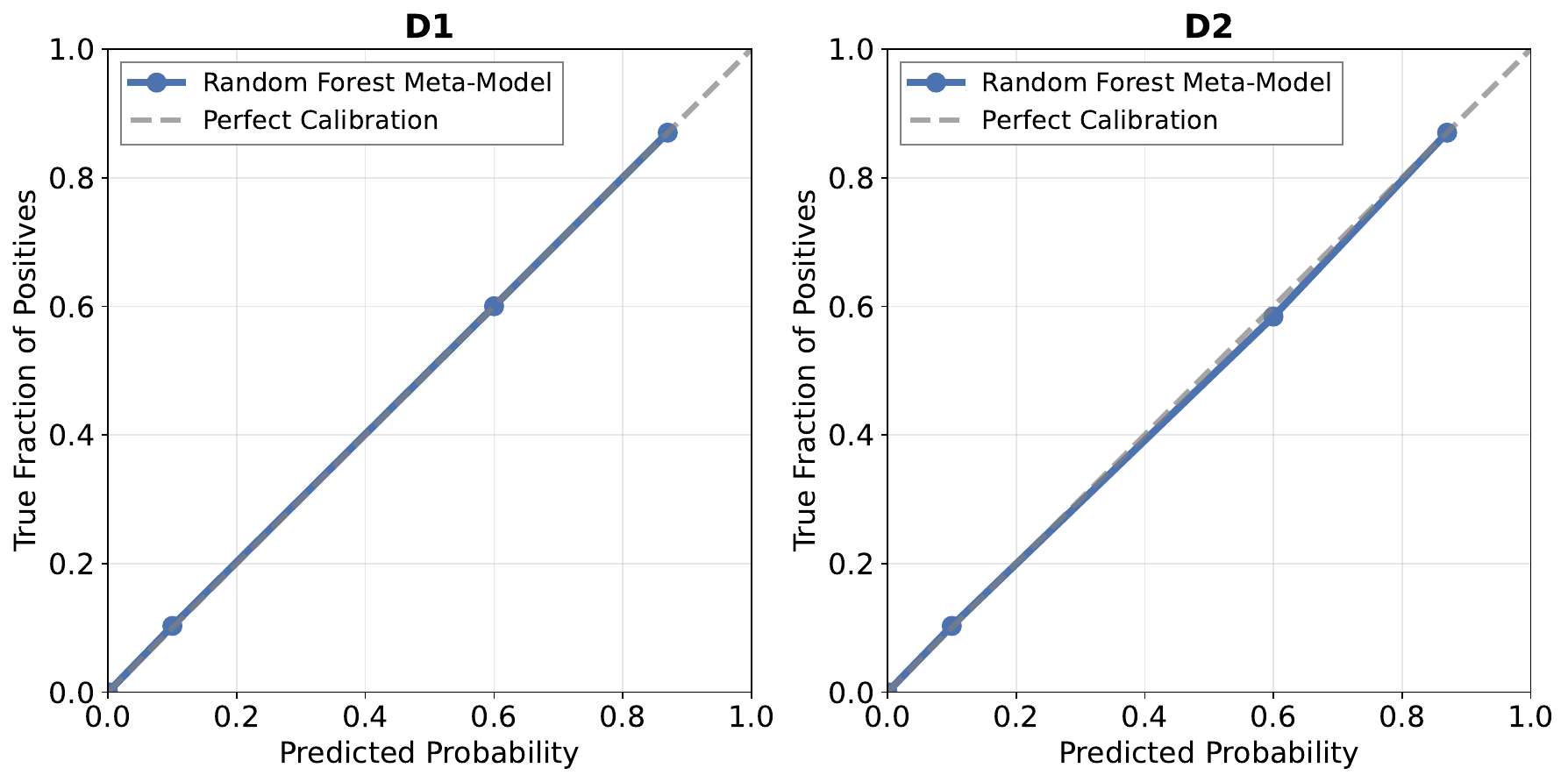}
   \caption{Reliability diagrams of the MQE framework on D1 and D2. In both datasets, the calibration curves remain close to the ideal diagonal, indicating well-calibrated probability estimates.}
    \label{fig:reliability_cruves}
\end{figure}

To evaluate the reliability of the probability estimates produced by MQE, we examine both calibration quality and predictive confidence. Table~\ref{tab:bs_ece} reports the Brier Score and Expected Calibration Error (ECE) for both datasets. In both cases, the framework achieves low Brier Scores and very small ECE values, indicating that the predicted probabilities are well aligned with the observed outcomes. This behavior is further illustrated in the reliability diagrams in Fig.~\ref{fig:reliability_cruves}, which are constructed using 15 probability bins. For both D1 and D2, the calibration curves remain close to the ideal diagonal, showing that the Random Forest meta-model provides confidence estimates that are well matched to the true event frequencies.

\begin{table}[htpb]
\centering
\caption{Brier Score and ECE}
\label{tab:bs_ece}
\begin{tabular}{lcc}
\toprule
 & Brier Score (\%)& ECE (\%)\\
\midrule
D1 & 6.07& 0.16\\
D2 & 0.57& 0.04\\
\bottomrule
\end{tabular}
\end{table}

The results indicate that MQE is not only effective as a classifier, but also reliable as a probabilistic decision model. This is particularly important in intrusion detection, where downstream actions such as threshold selection, alert prioritization, and automated response depend not only on the predicted class, but also on the confidence associated with that prediction. Well-calibrated probabilities therefore improve the practical usefulness of the framework in operational settings.

\subsection{Comparative Analysis}

\subsubsection{Internal Comparison}

Table~\ref{tab:ablation} compares three model groups under the same preprocessing and evaluation pipeline: classical baselines, adapted standalone quantum models, and the proposed MQE fusion variants. The classical models serve as strong tabular IDS references. The adapted quantum models are conceptually based on the cited QML methodologies, rather than exact reproductions, and are adjusted to the preprocessing, feature representation, binary IDS task, and simulation constraints of this study. The MQE rows correspond to the proposed fusion configurations.

\begin{table*}[htbp]
\centering
\caption{Internal comparison across classical baselines, adapted standalone quantum models, and proposed MQE fusion variants on D1 and D2. The adapted quantum models are conceptually based on the corresponding methodological references and are not intended as exact reproductions of the original implementations.}
\label{tab:ablation}
\resizebox{\textwidth}{!}{%
\begin{tabular}{llcccccc}
\toprule
\textbf{Dataset} & \textbf{Model} & \textbf{Acc. (\%)} & \textbf{F1 (\%)} & \textbf{AUPRC (\%)} & \textbf{ROC AUC} & \textbf{TPR@0.1\%FPR (\%)} & \textbf{TPR@1\%FPR (\%)} \\
\midrule

\multirow{12}{*}{D1}
& \multicolumn{7}{l}{\textit{Classical baselines}} \\
& RandomForest~\cite{RandomForest} & 99.73 & 98.29 & 99.80 & 0.9998 & 97.24 & 99.08 \\
& XGBoost~\cite{XGBoost} & 99.78 & 98.61 & 99.73 & 0.9996 & 98.16 & 99.08 \\
& LogReg~\cite{ScikitLearn} & 95.50 & 63.27 & 66.80 & 0.9178 & 46.01 & 49.08 \\
\cmidrule(lr){2-8}
& \multicolumn{7}{l}{\textit{Adapted standalone quantum models}} \\
& QNN\_Base & 97.10 & 77.86 & 89.47 & 0.9781 & 47.44 & 80.77 \\
& QNN\_Reupload~\cite{QNN_Reupload} & 97.20 & 79.10 & 85.18 & 0.9629 & 38.46 & 71.79 \\
& VQC\_AltAnsatz~\cite{VQC_AltAnsatz} & 94.60 & 50.00 & 70.00 & 0.9338 & 8.97 & 48.72 \\
& QSVM & 95.10 & 62.60 & 68.59 & 0.9130 & 15.38 & 47.44 \\
& QKNN~\cite{zardini2024quantum} & 96.50 & 63.16 & 65.23 & 0.8603 & 46.15 & 46.15 \\
\cmidrule(lr){2-8}
& \multicolumn{7}{l}{\textit{Proposed MQE fusion variants}} \\
& MQE\_Probability & 97.14 & 81.48 & 86.23 & 0.9674 & 40.74 & 70.37 \\
& MQE\_Margin & 96.86 & 78.43 & 83.93 & 0.9442 & 44.44 & 74.07 \\
& MQE\_Hard & 96.29 & 69.77 & 62.86 & 0.8122 & 35.50 & 62.99 \\

\midrule

\multirow{12}{*}{D2}
& \multicolumn{7}{l}{\textit{Classical baselines}} \\
& RandomForest~\cite{RandomForest} & 99.97 & 99.98 & $\sim$100 & 1.0000 & 99.99 & 100.00 \\
& XGBoost~\cite{XGBoost} & 99.96 & 99.97 & $\sim$100 & 1.0000 & 99.98 & 100.00 \\
& LogReg~\cite{ScikitLearn} & 95.24 & 96.90 & 95.92 & 0.9554 & 0.29 & 0.59 \\
\cmidrule(lr){2-8}
& \multicolumn{7}{l}{\textit{Adapted standalone quantum models}} \\
& QNN\_Reupload~\cite{QNN_Reupload} & 99.40 & 99.60 & 99.98 & 0.9994 & 88.02 & 99.87 \\
& QNN\_Base & 98.70 & 99.13 & 99.82 & 0.9952 & 59.62 & 89.10 \\
& QKNN~\cite{zardini2024quantum} & 98.50 & 98.99 & 99.19 & 0.9876 & 4.60 & 46.03 \\
& QSVM & 97.00 & 98.00 & 99.20 & 0.9916 & 0.54 & 93.81 \\
& VQC\_AltAnsatz~\cite{VQC_AltAnsatz} & 93.00 & 95.42 & 98.49 & 0.9680 & 17.36 & 32.57 \\
\cmidrule(lr){2-8}
& \multicolumn{7}{l}{\textit{Proposed MQE fusion variants}} \\
& MQE\_Hard & 99.14 & 99.43 & 99.61 & 0.9943 & 8.93 & 89.31 \\
& MQE\_Probability & 98.57 & 99.05 & 99.55 & 0.9933 & 7.89 & 78.92 \\
& MQE\_Margin & 98.57 & 99.04 & 99.54 & 0.9932 & 7.75 & 77.54 \\

\bottomrule
\end{tabular}
}
\end{table*}

The classical baselines achieve the strongest absolute performance on both datasets. Random Forest and XGBoost reach near-perfect results, especially on D2, confirming that tree-based ensemble models remain highly effective for tabular IDS data. Therefore, the purpose of MQE is not to replace these mature classical baselines, but to assess whether fusing heterogeneous quantum learners can improve the reliability of a quantum-classical IDS pipeline.

Among the adapted standalone quantum models, QNN-based variants generally perform better than QSVM, QKNN, and VQC\_AltAnsatz. On D1, QNN\_Reupload achieves an F1-score of 79.10\%, while QNN\_Base reaches 77.86\%. On D2, QNN\_Reupload gives the strongest standalone quantum result, with an F1-score of 99.60\% and a TPR@1\%FPR of 99.87\%. This suggests that data re-uploading increases the expressive capacity of the quantum classifier, especially when the dataset has clearer class separation.

The MQE variants show that fusion can improve the quantum pipeline, but the best fusion scheme depends on the dataset and metric. On D1, MQE\_Probability achieves the highest MQE F1-score, improving over both QSVM and QNN\_Base. MQE\_Margin gives the strongest low-FPR sensitivity among the MQE variants, reaching 74.07\% TPR@1\%FPR. On D2, MQE\_Hard performs best among the fusion schemes, reaching 99.43\% F1-score and 89.31\% TPR@1\%FPR. These results indicate that the fusion layer can extract useful information from the distinct behavior of the quantum learners, but its benefit depends on the form of the meta-features and the structure of the dataset.

The internal comparison supports three main observations. First, classical tree ensembles remain the strongest absolute baselines for these tabular IDS datasets. Second, QNN-based models provide stronger standalone quantum performance than the other adapted quantum baselines in most settings. Third, the proposed MQE fusion improves the quantum branch performance in several cases, particularly on D1, where individual quantum models show larger performance gaps.

\subsubsection{Comparison with Existing QML-IDS Models}

Table~\ref{tab:qml_ids_comparison} places MQE in relation to recent QML-based IDS studies. This comparison is intended only as external context, since the reported F1-scores are not directly comparable across studies due to differences in datasets, preprocessing, attack distributions, class definitions, and evaluation protocols.

\begin{table*}[htbp]
\centering
\caption{High-level comparison with existing QML-based IDS models.}
\label{tab:qml_ids_comparison}
\begin{tabular}{llcc}
\toprule
\textbf{Model} & \textbf{Dataset(s)} & \textbf{Setting} & \textbf{F1 (\%)} \\
\midrule
QuIDS~\cite{kumar2025quids} 
& Edge-IIoTset, ACI-IoT 
& IoT IDS 
& 86.40 \\

QML-IDS QCNN~\cite{abreu2025quantumnetsec} 
& TON IoT 
& Binary IDS 
& 98.16 \\

QSVM-IGWO~\cite{elsedimy2024novel} 
& BoT-IoT 
& IoT IDS 
& 97.78 \\

QSVM~\cite{said2023quantum} 
& CIC-DDoS2019 
& Binary IDS 
& 98.00 \\

QAE-QkNN~\cite{hdaib2024quantum} 
& CICIoT2023
& Binary IDS 
& 98.26 \\

QGAN-IDS~\cite{cirillo2025intrusion} 
& NSL-KDD 
& Anomaly IDS 
& 93.84 \\

\textbf{Ours (MQE\_Hard)} 
& \textbf{TON IoT} 
& \textbf{Binary IDS} 
& \textbf{99.43} \\
\bottomrule
\end{tabular}
\end{table*}

Compared with the reported QML-based IDS results, MQE achieves a higher F1-score on TON IoT. However, this external comparison should be interpreted cautiously because the studies use different datasets and protocols. The main evidence for the proposed framework remains the controlled internal comparison in Table~\ref{tab:ablation}, where all adapted baselines and MQE variants are evaluated under the same preprocessing and evaluation pipeline.


\subsection{Discussion}

The results provide several insights into the behavior of the proposed MQE framework. First, the analysis confirms that QSVM and QNN do not behave identically, even though both operate on the same preprocessed feature representation. The QSVM branch provides a kernel-based geometric decision function, while the QNN branch learns a trainable nonlinear mapping through its variational circuit. The error-overlap analysis shows that the two learners make different mistakes in several cases, which supports the motivation for combining them through a meta-learning layer rather than selecting only one quantum model.

Second, the fusion results show that the benefit of MQE depends on both the dataset and the form of the meta-features. On D1, probability-based fusion provides the best F1-score among the MQE variants, while margin-based fusion achieves the highest TPR at 1\% FPR. This indicates that confidence-related information is useful when the dataset is more difficult and the attack class is harder to separate. On D2, hard fusion performs best, suggesting that the individual quantum learners already produce reliable class decisions and that discrete outputs are sufficient for effective meta-learning. These findings show that the proposed fusion layer is not merely averaging predictions, but extracting useful information from the different decision behaviors of the quantum branches.

Third, the calibration and noise analyses provide additional evidence about the reliability of the framework. The low Brier Score and ECE values indicate that MQE produces well-calibrated probability estimates, which is important for IDS applications where alerts are often ranked or thresholded based on confidence. The noise robustness analysis further shows that the model maintains reasonable performance under moderate simulated noise, while performance degrades under stronger noise levels, as expected for NISQ-oriented quantum models. This suggests that future hardware-oriented implementations should consider noise-aware training, error mitigation, or circuit designs with improved tolerance to stochastic perturbations.

The comparison with classical baselines also clarifies the position of MQE. Random Forest and XGBoost remain the strongest absolute performers on these tabular IDS datasets, which is expected given the maturity of tree-based ensemble methods for structured data. Therefore, the proposed framework should not be interpreted as replacing classical IDS models. Instead, its contribution is to show that heterogeneous quantum learners can be organized into a more reliable system-level pipeline, improving over several standalone quantum branches and offering a structured way to study quantum-model fusion for intrusion detection.

Our MQE demonstrates that meta-learning can strengthen quantum-classical IDS pipelines by exploiting disagreement, confidence patterns, and dataset-specific behavior across quantum learners. The strongest evidence for this contribution comes from the controlled internal comparison, where all adapted quantum baselines and MQE variants are evaluated under the same preprocessing and metric protocol. These results position MQE as a useful step toward more robust and modular QML-based IDS frameworks, while also highlighting the need for broader evaluation under realistic traffic drift, larger-scale datasets, and hardware-aware execution.

\section{Conclusion}
\label{sec:conclusion}

This work investigated whether heterogeneous quantum learners can be combined to improve the reliability of QML-based intrusion detection. We introduced MQE, a hybrid quantum--classical framework that integrates a QSVM and a QNN through a Random Forest meta-learner. The goal was not to replace mature classical IDS models, but to assess whether meta-level fusion can strengthen the performance and robustness of quantum learning components under challenging IDS conditions.

Experiments on CICIDS2017 and TON~IoT show that MQE improves over several standalone quantum branches, with gains in F1-score, AUPRC, and true-positive rate under low false-positive rate constraints. The disagreement and error-overlap analyses further show that QSVM and QNN make different types of mistakes, allowing the meta-learner to exploit their distinct decision behaviors. The fusion results also reveal that the most effective meta-feature representation depends on the dataset: probability and margin features are more useful on the harder CICIDS2017 setting, while hard-label fusion performs best on TON~IoT.

The calibration and noise analyses provide additional evidence of the framework's reliability. MQE produces well-calibrated probability estimates, which is important for threshold-based IDS decision-making, and maintains reasonable performance under moderate simulated quantum noise. However, performance degrades under stronger noise levels, indicating that future hardware-oriented implementations should consider noise-aware training, error mitigation, and more robust circuit designs.

Overall, MQE should be interpreted not as a universally dominant IDS classifier, but as evidence that structured meta-fusion can strengthen QML-based IDS pipelines. While classical ensembles remain stronger absolute baselines and no single MQE variant dominates all metrics across both datasets, the framework demonstrates that combining heterogeneous quantum learners can improve selected operationally relevant metrics and provide a modular foundation for future quantum-classical IDS research. Future work will extend this framework to multi-class attack detection, traffic drift scenarios, larger-scale benchmarks, and hardware-aware deployment.

\section*{Acknowledgment}
 This work was supported in part by the NYUAD Center for Quantum and Topological Systems (CQTS), funded by Tamkeen under the NYUAD Research Institute grant CG008.
\bibliographystyle{IEEEtran}
\bibliography{refs}

\end{document}